 \documentclass[final,5p,times,twocolumn]{elsarticle}

\usepackage{amssymb}
\usepackage{amsmath, amsthm, bm, braket}

\usepackage{lineno}
\usepackage{color}

\newcommand{\bi}[1]{\ensuremath{\boldsymbol{#1}}}

\newcommand{\abs}[1]{\ensuremath{\left| #1 \right|}}
\newcommand{\tomo}[1]{\textcolor[rgb]{0.0, 0.0, 0.0}{#1}}

\def \beq{\begin{equation}}
\def \eeq{\end{equation}}
\def \beqa{\begin{eqnarray}}
\def \eeqa{\end{eqnarray}}
\def \bir{\bi{r}}

\def \bip{\bi{p}}

\def \twop{$2p$}

\journal{Nuclear Physics B}

\begin{document}

\begin{frontmatter}

\title{Two-proton emission as source of spin-entangled proton pairs}

\author[ibakou,riken]{Tomohiro Oishi}
\author[riken,toudai]{Masaaki Kimura}

\affiliation[ibakou]{organization={Ibaraki College in National Institute of Technology (KOSEN)},
            addressline={866 Nakane},
            city={Hitachinaka},
            postcode={312-8508}, 
            state={Ibaraki},
            country={Japan}}
\affiliation[riken]{organization={RIKEN Nishina Center for Accelerator-Based Science},
            addressline={2-1 Hirosawa},
            city={Wako},
            postcode={351-0198}, 
            state={Saitama},
            country={Japan}}
\affiliation[toudai]{organization={Department of Physics / Quark Nuclear Science Institute, Graduate School of Science, The University of Tokyo},
            addressline={7-3-1 Hongou},
            city={Bunkyou},
            postcode={113-0033},
            state={Tokyo},
            country={Japan}}

\begin{abstract}
We show that a two-proton emitter with a diproton-correlated initial state can act as a source of spin-correlated proton pairs.
Using a time-dependent three-body model, we investigate the two-proton emission of $^{16}$Ne ($^{14}$O$+2p$) and analyze the spin correlation of the emitted protons.
We find that, when the emission proceeds as a democratic three-body process from an initial state containing a spin-singlet diproton correlation,
the emitted protons exhibit a pronounced spin-correlation pattern exceeding the local-hidden-variable bound.
This spin correlation closely resembles that of a pure spin-singlet pair.
In contrast, this pattern is lost when the process is dominated by the sequential emission or when the initial diproton correlation is absent.
These results demonstrate that a certain class of two-proton emitters can deliver spin-entangled proton pairs, and their spin correlation reflects the diproton correlation embedded in the initial state.
\end{abstract}



\begin{keyword}
Entanglement, Two-proton emission, Time-dependent calculation


\end{keyword}

\end{frontmatter}

%%\linenumbers

%\section{}
%\label{}

%%%%%%%%%%%%%%%%%%%%%%%%%%%%%%%%%%%%%%%%%%%%%%%%%%%%%%%%%%%%%%%%%%%%%%%%%%%%%%%%
%%%%\input{POWDAM_PLB_Sections.tex}
%%%%%%%%%%%%%%%%%%%%%%%%%%%%%%%%%%%%%%%%%%%%%%%%%%%%%%%%%%%%%%%%%%%%%%%%%%%%%%%%

\section{Introduction}
Spin correlation is a direct manifestation of quantum entanglement \cite{04Bell,64Bell,1969CHSH}.
Violation of the local-hidden-variable (LHV) bound in terms of Clauser-Horne-Shimony-Holt (CHSH) formulation has been certified in atomic and optical systems~\cite{1982Aspect,1998Weihs,2001Rowe,2015Giustina,2015Shalm}.
Nuclear systems offer a conceptually distinct arena, where correlated fermions emerge naturally with strong interactions in a finite many-body environment~\cite{2006Sakai,1976Rachti,2004Polachic}.
A pioneering experiment by Sakai Sakai {\it et al.} measured the spin correlation of two protons emitted in the reaction $^2\mathrm{H}(p,{}^2\mathrm{He})n$.
They observed a clear violation of the Bell-CHSH inequality \cite{2006Sakai}, demonstrating the generation of spin-entangled proton pairs.

Two-proton (\twop) decay of proton-rich nuclei provides a natural setting for producing correlated proton pairs~\cite{02Gri_15,02Pfu,2012Pfu_rev,2009Gri_rev,2016Golubkova,08Blank_01,08Blank_02,2023Pfutzner_rev,2019Qi_rev}.
In so-called democratic \twop~emitters \cite{2016Golubkova,2023Pfutzner_rev}, where a sequential emission is suppressed, two valence protons are confined inside the Coulomb barrier for a finite lifetime and interact within a finite spatial volume.
The proton-proton interaction can compete with, or even dominate over, the interaction between each proton and the core (daughter) nucleus.
This interaction particularly enhances the spin-singlet component and thereby gives rise to a diproton correlation~\cite{2003Dean,05BB,2006Matsuo,07Hagi_01,2010Oishi,2014Oishi,2017Oishi,2025Oishi_16Ne,2025Oishi_QUBITE,2021Wang_Naza}.
Once the decay sets in, however, the two protons tunnel through the long-range Coulomb field and evolve under three-body dynamics~\cite{08Blank_01,2009Gri_rev}.
Thus, measured spatial and kinematic correlations are governed primarily by this dynamics and retain a limited sensitivity to the initial structure~\cite{2014Brown,2016Charity_EPJCON}.
This behavior is manifested differently in \twop~emitters such as $^6$Be, $^{16}$Ne, and $^{45}$Fe~\cite{2012Pfu_rev,2023Pfutzner_rev}.
In particular, for $^{16}$Ne, both experimental and theoretical studies indicate that spatial and kinematic correlations provide only limited information on the initial configuration~\cite{2014Brown,12Gri},
making it a stringent test case for identifying robust observables against three-body dynamics.
This generic loss of sensitivity highlights the need to explore whether the \twop~decay can generate and preserve nontrivial quantum correlations.
Because the Coulomb interaction governing the three-body dynamics is essentially spin independent, spin correlations are expected as robust and can provide an access to the initial structure.

In this Letter, we study the \twop~decay of $^{16}$Ne to examine whether a \twop~emitter with the initial diproton correlation can act as a source of spin-correlated proton pairs.
For this purpose, we employ a time-dependent three-body model of $^{14}$O$+2p$ \cite{2025Oishi_16Ne}.
This model enables us to monitor how the \twop-spin correlation is generated and preserved during the decaying process \cite{2025Oishi_QUBITE}.

%%%%%%%%%%%%%%%%%%%%%%%%%%%%%%%%%%%%%%%%%%%%%%%%%%%%%%%%%%%%%%%%%%%%%%%%%%%%%%%%%%%%%%%%%
{\it Spin correlation.}
To quantify the spin correlation between the emitted protons, we employ the CHSH formulation \cite{1969CHSH,2006Sakai}.
The \twop-spin correlation is defined as
\beq
S(\Phi) = \max  \left\{  \abs{S_{1}(\Phi)},  \abs{S_{2}(\Phi)}, \abs{S_{3}(\Phi)},  \abs{S_{4}(\Phi)}  \right\},
\eeq
where
\beqa
S_{1} (\Phi) &=& -\Braket{A_1 B_1}+\Braket{A_2 B_1}+\Braket{A_1 B_2}+\Braket{A_2 B_2}, \nonumber \\
S_{2} (\Phi) &=&  \Braket{A_1 B_1}-\Braket{A_2 B_1}+\Braket{A_1 B_2}+\Braket{A_2 B_2}, \nonumber \\
S_{3} (\Phi) &=&  \Braket{A_1 B_1}+\Braket{A_2 B_1}-\Braket{A_1 B_2}+\Braket{A_2 B_2}, \nonumber \\
S_{4} (\Phi) &=&  \Braket{A_1 B_1}+\Braket{A_2 B_1}+\Braket{A_1 B_2}-\Braket{A_2 B_2}. \label{eq:CHSHs}
\eeqa
For an arbitrary two-fermion state $\ket{\Psi(1,2)}$, these expectation values for the two observers conventionally named as ``Alice'' and ``Bob'' are determined as
\beq
\Braket{A_i B_j} = \Braket{ \Psi(1,2)  \mid \hat{A}_{i,\Phi} (1) \otimes \hat{B}_{j,\Phi}(2) \mid  \Psi(1,2)}.
\eeq
Here $\hat{A}_{i,\Phi} (1)$ and $\hat{B}_{j,\Phi} (2)$ are spin-projection operators for proton 1 and 2, respectively:
\beq
\begin{array}{l}
  \hat{A}_{1,\Phi} (1) = \hat{\sigma}_z(1), \\
  \hat{A}_{2,\Phi} (1) = \hat{\sigma}_z(1) \cos 2\Phi +\hat{\sigma}_x(1) \sin 2\Phi, \\
  \hat{B}_{1,\Phi} (2) = \hat{\sigma}_z(2) \cos \Phi +\hat{\sigma}_x(2) \sin \Phi, \\
  \hat{B}_{2,\Phi} (2) = \hat{\sigma}_z(2) \cos \Phi -\hat{\sigma}_x(2) \sin \Phi.
\end{array}
\eeq
Each operator projects the proton's spin onto the fixed direction specified by the orientation angle $\Phi$ \cite{1969CHSH,2006Sakai}.

%%%%%%%%%%%%%%%%%%%%%%%%%%%%%%%%%%%%%%%%%%%%%%%%%%%%%%%%%%%%%%%%%%%%%%
\begin{figure}[t] \begin{center}
\includegraphics[width = \hsize]{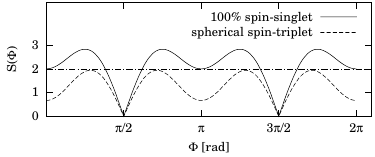}
\caption{Spin correlation of the pure spin-singlet ($S_{12}=0$) and spherical spin-triplet ($S_{12}=1$) states.
The limit of LHV theory is given as $S(\Phi) =2$ \cite{1969CHSH}.}
\label{fig:PEBIRE}
\end{center} \end{figure}
%%%%%%%%%%%%%%%%%%%%%%%%%%%%%%%%%%%%%%%%%%%%%%%%%%%%%%%%%%%%%%%%%%%%%%

Figure \ref{fig:PEBIRE} illustrates the characteristic behavior of $S(\Phi)$.
For a pure spin-singlet ($S_{12}=0$) state, $S (\Phi)$ reaches the maximal value $2\sqrt{2}$ at specific angles $\Phi$, signaling a maximal violation of the Bell-CHSH inequality (Tsirelson's bound~\cite{1980Tsirelson}).
A similar pattern was experimentally observed by Sakai {\it et al.}~\cite{2006Sakai}.
In contrast, for a spherical spin-triplet ($S_{12}=1$) state, where the magnetic substates are equally mixed, $S(\Phi)$ remains below $2$.

%%%%%%%%%%%%%%%%%%%%%%%%%%%%%%%%%%%%%%%%%%%%%%%%%%%%%%%%%%%%%%%%%%%%%%%%%%%
\begin{table}[b] \begin{center}
\caption{Parameters for potentials.
Note that the democratic and symmetric cases have the same set of parameters.
}
\label{table:KIM}
\catcode`? = \active \def?{\phantom{0}} %define `?' as ' '(one-blank).
  \begingroup \renewcommand{\arraystretch}{1.2}
  \begin{tabular*}{\hsize} { @{\extracolsep{\fill}} llcc}
\hline \hline%%%%%%%%\multicolumn{3}{c}{(even-even)}
          & &Democratic \&   &Sequential \\
          & &Symmetric          &           \\  \hline
 ~$V_{WS}$   &$r_0$~[fm]         &$1.23$   &$1.23$   \\
               &$a_0$~[fm]         &$0.82$   &$0.82$   \\
               &$V_{0}$~[MeV]      &$-42.25$    &$-48.93$   \\
               &$U_{ls}$~[MeV$\cdot$fm$^2$]  &$54.1$  &$54.1$   \\
~ $v_{pp}$ &$w_0$~[MeV$\cdot$fm$^3$]  &$-527$  &$-527$   \\
         &$f_{pp}$  &$1.0$  &$0.2$   \\
\hline \hline
  \end{tabular*}
  \endgroup
  \catcode`? = 12 %initialize `?'.
\end{center} \end{table}
%%%%%%%%%%%%%%%%%%%%%%%%%%%%%%%%%%%%%%%%%%%%%%%%%%%%%%%%%%%%%%%%%%%%%%%%%%%

\section{Three-Body Model}
We describe the \twop~decay of $^{16}$Ne with a time-dependent three-body model consisting of the $^{14}$O core and two valence protons \cite{2025Oishi_16Ne,2025Oishi_QUBITE,2021Wang_Naza}.
\tomo{The first-excited state of $^{14}$O is at $5.17$ MeV \cite{NNDC_Chart}, which is higher than the experimental \twop-emission energy of $^{16}$Ne, $Q_{2p} =1.4$ MeV \cite{2021Wang_AME,02Gri,08Muk,10Muk}.
Thus, the effect of core excitation is expected as small, and neglected in this Letter.}

The three-body Hamiltonian is given as \cite{2025Oishi_16Ne}
\beq
\hat{H}_{3B} = \hat{h}(\bir_1)+\hat{h}(\bir_2) +v_{pp}(\bir_1,\bir_2) +\frac{\bip_1 \cdot \bip_2}{m_C}. \label{eq:H3Body}
\eeq
Here $\hat{h}(\bir_i)$ includes the Woods-Saxon and Coulomb potentials to describe the core-proton subsystem:
\beq
\hat{h}(r_i) = -\frac{\hbar^2}{2\mu} \frac{d^2}{dr_i^2} + \frac{\hbar^2}{2\mu} \frac{l(l+1)}{r_i^2} +V_{WS}(r_i) +V_{Coul}(r_i),
\eeq
with $\mu = m_p m_C /(m_p + m_C)$, where $m_C$ and $m_p$ are the core and proton masses, respectively.
\tomo{These potentials are determined as
\beqa
&& V_{WS}(r) = \left[V_0 +U_{ls} (\bi{l} \cdot \bi{s}) \frac{1}{r} \frac{d}{dr} \right]  \left( \frac{1}{1 + e^{(r-R_0)/a_0}} \right), \nonumber  \\
&&
V_{Coul}(r)= \left\{ \begin{array}{ll}
  \displaystyle  \alpha \hbar c \frac{Z}{r}                                                   &(r \ge R_0)  \nonumber  \\
  \displaystyle  \alpha \hbar c \frac{Z}{2R_0} \left[ 3-\left(\frac{r}{R_0} \right)^2 \right]  &(r<R_0) \end{array} \right.   , \label{eq:FWS15}
\eeqa
where $R_{0} = r_{0} \cdot 14^{1/3}$, $\alpha  =\frac{e^2}{4\pi \epsilon_0 \hbar c} =\frac{1}{137.036}$, and $Z=8$.
Other parameters are summarized in TABLE \ref{table:KIM}.
On the other side, the proton-proton interaction includes the vacuum and additional terms \cite{2025Oishi_16Ne}:
\beq
v_{pp}(\bm r_1,\bm r_2)= v_{pp,\mathrm{vac}}(\bm r_1,\bm r_2)+ v_{pp,\mathrm{add}}(\bm r_1,\bm r_2). \label{eq:vpp}
\eeq
These terms read
\beqa
&& v_{pp,{\rm vac}}(\bir_1,\bir_2) = f_{pp} \cdot \Bigl[ V_R e^{-a_R d^2} +V_S e^{-a_S d^2} \hat{P}_{S=0}  \Bigr. \nonumber  \\
&& ~~~~~~~~~~~~~~~~~~~~~~~~~~\Bigl. +V_T e^{-a_T d^2} \hat{P}_{S=1}  \Bigr]
  +\frac{\alpha  \hbar c}{d},  \label{eq:vpptue}  \\
&&
v_{pp,{\rm add}} (\bir_1,\bir_2) = f_{pp} \cdot w_0 e^{-(R-R_0)^2 /B_0^2} \delta (\bir_1-\bir_2), \label{eq:vppadd}
\eeqa
where $d= \abs{\bir_2 -\bir_1}$ and $R=\abs{(\bir_1 +\bir_2) /2}$.
The operators $\hat{P}_{S=0}$ and $\hat{P}_{S=1}$ are the projectors to the spin-singlet and spin-triplet channels, respectively.
Parameters for $v_{pp,{\rm vac}}$ were given as
$V_R=200$ MeV,
$V_S=-91.85$ MeV,
$V_T=-178$ MeV,
$a_R=1.487$ fm$^{-2}$,
$a_S=0.465$ fm$^{-2}$, and $a_T=0.639$ fm$^{-2}$, in order to reproduce the experimental scattering data of two protons \cite{77Thom}.
For the additional term, $R_0 = r_0 \cdot 14^{1/3}$ and $B_0 = 0.6 R_0$.
This term accounts for medium effects to reproduce the experimental three-body energy, $Q_{2p} =1.4$ MeV, of $^{16}$Ne \cite{2021Wang_AME,02Gri,08Muk,10Muk}.
Notice that $v_{pp,{\rm add}} (\bir_1,\bir_2) = 0$ when $d \longrightarrow \infty$ or $R \longrightarrow \infty$.
Thus, it does not affect the properties of two-body subsystems in vacuum.
We also use the tuning factor $f_{pp}$ for the nuclear-force parts: see TABLE \ref{table:KIM}.
Reason for employing these parameters is explained later with the calculated results and discussions.
}

To construct the model space for the \twop~state with $J^\pi=0^+$, we first generate a set of single-particle states by solving the core-proton subsystem $\hat{h} (r)$ in a radial box of $r_{\rm max}=80$ fm.
Continuum states are thus discretized.
The \twop~state is then expanded with the anti-symmetrized products of single-particle states.
To keep the model space tractable, cutoffs are introduced for the single-particle orbital angular momentum and energy, $l \le 7$ and $E \le 24$ MeV.

%%%%%%%%%%%%%%%%%%%%%%%%%%%%%%%%%%%%%%%%%%%%%%%%%%%%%%%%%%%%%%%%%%%%%%
\begin{figure}[t] \begin{center}
\includegraphics[width = \hsize]{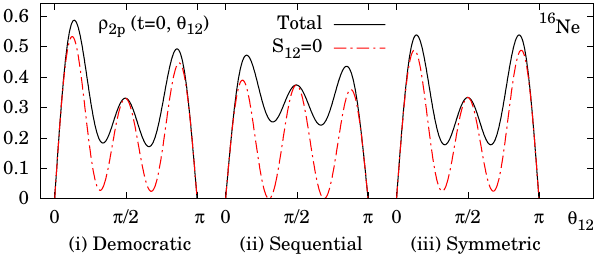}
\caption{Probability distributions of the initial \twop~state in $^{16}$Ne as a function of the opening angle $\theta_{12}$.
These distributions are obtained by integrating over the radial coordinates of two protons.
Results are shown for (i) the Democratic, (ii) Sequential, and (iii) Symmetric cases.
}
\label{fig:initial_state}%%%\label{fig:21671}
\end{center} \end{figure}
%%%%%%%%%%%%%%%%%%%%%%%%%%%%%%%%%%%%%%%%%%%%%%%%%%%%%%%%%%%%%%%%%%%%%%

Time evolution of \twop~state is calculated as $\ket{\Psi(t)} = \exp ( -it \hat{H}_{3B} / \hbar ) \ket{\Psi(0)}$,
where the initial state is solved as the confined \twop~state inside the Coulomb barrier \cite{2025Oishi_16Ne}.
Then, to extract the emitted component, we define the decaying state as
\beq
\ket{\Psi_d(t)} = \ket{\Psi(t)}  -\beta(t) \ket{\Psi(0)},
\eeq
where $\beta(t)$ is the survival coefficient, $\beta(t)=\Braket{\Psi(0) | \Psi(t)}$.
The spin correlation $S(\Phi)$ is evaluated for this decaying state $\ket{\Psi_d(t)}$ by integrating out the coordinate degrees of freedom and retaining only the spin components.
In practice, this is achieved by projecting $\ket{\Psi_d(t)}$ onto the $S_{12}=0$ and $S_{12}=1$ channels \cite{2025Oishi_QUBITE}.

\section{Results and Discussions}
To clarify how the spin correlation $S(\Phi)$ reflects the underlying decay dynamics,
we compare (i) a democratic-emission case and (ii) a sequential-emission case.
Then, we later discuss (iii) a symmetric case, where the initial state of (i) is modified so as to lack the diproton correlation.

In (i) the democratic-emission case \cite{2025Oishi_16Ne}, the core-proton Hamiltonian $\hat{h}(r_i)$ was tuned to reproduce the observed one-proton resonances, $1.5$ and $2.8$ MeV in the $s_{1/2}$ and $d_{5/2}$ channels, respectively, of $^{15}$F \cite{2003Peters}.
Practically, we use the same parameters in the ``prompt'' case in Ref. \cite{2025Oishi_16Ne}.
The additional proton-proton interaction, $v_{pp,\mathrm{add}}$, was optimized to reproduce the experimental \twop~energy of $^{16}$Ne, $Q_{2p} =1.4$ MeV \cite{2021Wang_AME,02Gri,08Muk,10Muk}.
With this setup, the initial state exhibits a diproton correlation, which manifests itself in Fig.~\ref{fig:initial_state}~(i) as an asymmetric probability distribution of the opening angle $\theta_{12}$ between two protons.
Note that the main-resonance width was obtained as $\Gamma_{2p} = 1.4 \times 10^{-4}$ MeV \cite{2025Oishi_16Ne}, corresponding to the mean lifetime, $\tau = 4.7 \times 10^{-18}$ seconds.

In (ii) the sequential-emission case, %the potential $V_{WS}(r_i)$ is modified from (i) such that the one-proton $s_{1/2}$ resonance is located around $1.1$ MeV.
the central term in the core-proton Woods-Saxon potential is multiplied by $1.158$.
Owing to this deeper $V_{WS}(r_i)$,
the nuclear part of $v_{pp}$ is correspondingly weakened by the factor $0.2$ to keep the same \twop~energy, $Q_{2p} =1.4$ MeV: \tomo{see TABLE \ref{table:KIM}.}
With this setting, a sequential emission is promoted \cite{2012Pfu_rev,2009Gri_rev,2016Golubkova}.
The mean lifetime in this case is evaluated as $\tau = 9.1 \times 10^{-20}$ seconds. %($\Gamma_{2p} \cong 7.2 \times 10^{-3}$ MeV).
In Fig.~\ref{fig:initial_state}~(ii), the initial state shows a weaker diproton correlation, as confirmed by the suppressed asymmetry in the opening-angle distribution.

%%%%%%%%%%%%%%%%%%%%%%%%%%%%%%%%%%%%%%%%%%%%%%%%%%%%%%%%%%%%%%%%%%%%%%
\begin{figure}[t] \begin{center}
\includegraphics[width = \hsize]{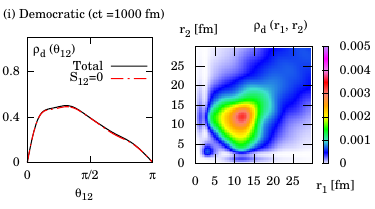}
\includegraphics[width = \hsize]{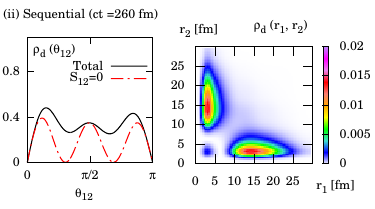}
\caption{Probability densities of the \twop-decaying state of $^{16}$Ne shown as functions of the opening angle $\theta_{12}$ (left panels) and of the radial coordinates $r_1= \abs{\bir_1}$ and $r_2 =\abs{\bir_2}$ (right panels).
Results are shown for (i) the Democratic (upper panels) and (ii) Sequential decay cases (lower panels).}
\label{fig:2p_emission}%%%%%\label{fig:77712}
\end{center} \end{figure}
%%%%%%%%%%%%%%%%%%%%%%%%%%%%%%%%%%%%%%%%%%%%%%%%%%%%%%%%%%%%%%%%%%%%%%

Figure \ref{fig:2p_emission} shows the probability density of the emitted protons at a certain time, $\rho_d(\bir_1,\bir_2, t)= \abs{\Psi_d(\bir_1,\bir_2, t)}^2$.
The time evolution of the cases (i) and (ii) indeed exhibits qualitatively different dynamics.
In case (i), the distribution is concentrated at small opening angles between the two protons, and the asymmetry is more pronounced than the initial state shown in Fig. \ref{fig:initial_state} (i).
We also find that the probability density is localized around $r_1 \simeq r_2$, indicating that two protons are emitted in a spatially correlated manner.
These features demonstrate that this decay proceeds as a genuine three-body process, in which the two protons are released simultaneously.
In case (ii), on the other side, the angular correlation is significantly weakened compared to case (i), and even slightly reduced relative to the initial distribution in Fig. \ref{fig:initial_state} (ii).
Moreover, the probability density is concentrated either $r_1 \simeq 0$ or $r_2 \simeq 0$, indicating that the two protons are emitted independently, one after the other.
Namely, a sequential emission through the intermediate core-proton resonance is simulated.

%%%%%%%%%%%%%%%%%%%%%%%%%%%%%%%%%%%%%%%%%%%%%%%%%%%%%%%%%%%%%%%%%%%%%%
\begin{figure}[t] \begin{center}
\includegraphics[width = \hsize]{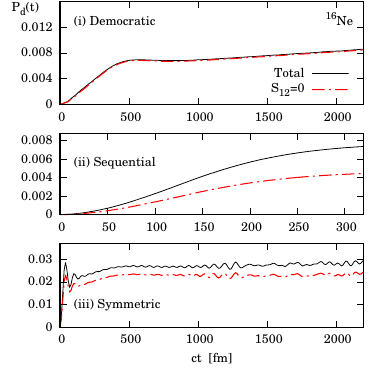}
\caption{
Time-dependent decaying probability $P_{d}(t)$ for (i) the Democratic, (ii) Sequential, and (iii) Symmetric cases.
Partial contributions of the $S_{12}=0$ channel are also plotted.
}
\label{fig:decay_prob}%%%%\label{fig:PT_DEC}
\end{center} \end{figure}
%%%%%%%%%%%%%%%%%%%%%%%%%%%%%%%%%%%%%%%%%%%%%%%%%%%%%%%%%%%%%%%%%%%%%%

A further notable difference between the two cases appears in the coupled-spin channels.
In case (i) of Fig. \ref{fig:2p_emission}, the \twop-decaying probability is entirely exhausted by the $S_{12}=0$ component, whereas the $S_{12}=1$ contribution is negligible.
In case (ii), by contrast, the $S_{12}=0$ and $S_{12}=1$ components are comparable.
Namely, this case (ii) considerably loses the strong spin correlation.

The coupled-spin properties observed at the specific time in Fig.~\ref{fig:2p_emission} persist throughout the decay process.
This is confirmed by Fig.~\ref{fig:decay_prob}, which shows the time dependence of the decaying probability, $P_{d}(t)=\braket{\Psi_d(t)|\Psi_d(t)}$.
At all times, the decaying state in case (i) remain dominated by the $S_{12}=0$ channel,
whereas, in case (ii), the decay always involves a mixture of $S_{12}=0$ and $S_{12}=1$ components.
It is worth noting that the two cases have different time scales, as seen in Fig.~\ref{fig:decay_prob}.
This indicates the different tunneling probabilities of the democratic and sequential processes.
Accordingly, the \twop-density distributions in Fig.~\ref{fig:2p_emission} are shown at times corresponding to the same probability: $P_{\rm decay} (t)= 0.68\%$ at $ct=1000$~fm for case (i) and $260$~fm for case (ii).

\begin{figure}[t] \begin{center}
\includegraphics[width = \hsize]{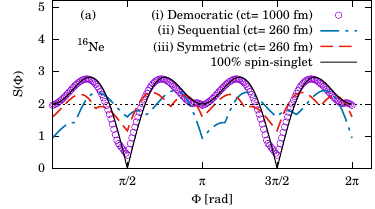}
\includegraphics[width = \hsize]{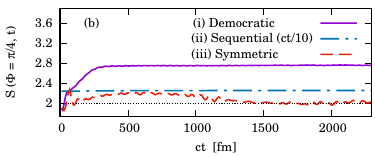}
\caption{(Top) Angular dependence of $S(\Phi)$ evaluated at representative times corresponding to the same decay probability $P_d \simeq 0.68\%$ for the (i) Democratic, (ii) Sequential, and (iii) Symmetric cases.
A pure spin-singlet case is also shown.
(Bottom) Spin correlations $S(\Phi = \pi /4)$ calculated for the time-dependent decaying states of $^{16}$Ne.
In the (ii) Sequential case, its time is scaled by $10$, due to the short lifetime.}
\label{fig:spin_correlation}%%%%%%%%%%%%%\label{fig:GEMUTO}
\end{center} \end{figure}

Contrast between the democratic and sequential emissions is clearly presented by the spin-correlation function $S(\Phi)$.
Figure~\ref{fig:spin_correlation} (a) shows that the case (i) exhibits a pattern of $S(\Phi)$ almost identical to that of a pure $S_{12}=0$ state,
where its maximal value reaches $\simeq 2\sqrt{2}$ exceeding the LHV bound.
Thus, a strong spin correlation in the initial state is preserved during the decay and manifests itself in the asymptotic observables.
In the case (ii), on the other side, the admixture of $S_{12}=0$ and $S_{12}=1$ components washes out this characteristic pattern.

Stability of the spin correlation is demonstrated in Fig.~\ref{fig:spin_correlation} (b), which shows the time evolution of $S(\Phi)$ at $\Phi= \pi/4$.
In both of cases (i) and (ii), their values of $S(\Phi)$ remain constant over long times, indicating that the spin-correlation pattern shown in Fig.~\ref{fig:spin_correlation} (a) is robust throughout the decaying process.
At later times, the emitted protons reach sufficiently large distances, where the subsequent dynamics is dominated by the spin-independent Coulomb interaction.
Consequently, we expect that the spin correlations are preserved and accessible even at macroscopic distances.
In this sense, case (i) realizes a femtometer-scale process, in which two protons are delivered to macroscopically separated locations keeping their spin entanglement.
Therefore, a measurement of $S(\Phi)$ downstream of the decay process, as realized in the experimental geometry of Ref.~\cite{2006Sakai}, should directly access the pattern shown in Fig.~\ref{fig:spin_correlation} (a).

%%%%%%%%%%%%%%%%%%%%%%%%%%%%%%%%%%%%%%%%%%%%%%%%%%%%%%%%%%%%%%%%%%%%%%
%%%%%%%%%%%%%%%%%%%%%%%%%%%%%%%%%%%%%%%%%%%%%%%%%%%%%%%%%%%%%%%%%%%%%%

Finally, we examine how the spin correlation depends on the presence of initial diproton correlation.
In (iii) the symmetric case, the initial wave function is constructed by modifying that of case (i).
Specifically, the odd-parity components responsible for the asymmetric opening-angle distribution are removed.
As result, the initial state without a diproton correlation is fabricated, being displayed as the symmetric distribution in Fig.~\ref{fig:initial_state}~(iii).
The subsequent time evolution is calculated with the same Hamiltonian in case (i).
The corresponding decay probability is shown in Fig.~\ref{fig:decay_prob}~(iii).
Because the initial wave function is modified by hand, high-energy components are admixed, and they are emitted within the first $\simeq 50$~fm/$c$.
After this early stage, the decay probability exhibits a stable behavior, apart from small residual oscillations. 
Note that the emitted \twop-wave function is still dominated by the $S_{12}=0$ channel, even though the initial state lacks a diproton correlation.
Namely, the Hamiltonian itself induces a certain degree of $S_{12}=0$ preference during the decay.

Remarkably, no significant structure appears in the spin correlation $S(\Phi)$: see the case (iii) in Fig.~\ref{fig:spin_correlation} (a).
Thus, a strong $S_{12} =0$ fraction alone is not sufficient to produce a nontrivial spin-correlation pattern.
Altogether, these results indicate that a pronounced spin correlation is observed only when the initial state contains a diproton correlation and the Hamiltonian suppresses the sequential decay [case (i)].
When the Hamiltonian allows sequential decay [case (ii)], or when the initial diproton correlation is absent despite the suppression of sequential decay [case (iii)], the spin-correlation pattern is lost.
Consequently, a democratic \twop~emitter with the initial diproton correlation can act as a source of spin-correlated proton pairs.

\section{Summary}
We investigated the spin correlation in the \twop~emission from $^{16}$Ne with a time-dependent three-body model, focusing on its sensitivity to the initial diproton correlation.
When the decay proceeds as a democratic three-body process from an initial state containing a localized $S_{12} =0$ configuration,
the spin correlation $S(\Phi)$ exhibits a characteristic angular dependence closely resembling that of a pure $S_{12} =0$ state.
In contrast, when the decay is dominated by a sequential emission, the admixture of $S_{12} =1$ components washes out this pattern.
Moreover, in the absence of an initial diproton correlation, a pronounced spin-correlation structure does not develop, even if the sequential decay is suppressed.
As conclusion, democratic \twop~emitters can serve as a source of spin-correlated proton pairs, and their spin correlation reflects the initial diproton correlation.

\tomo{The core nucleus $^{14}$O is treated as an inert core in this Letter.
For investigating whether the core excitation disrupts the spin correlation or not, a technical improvement is necessary.}

This Letter focuses solely on $^{16}$Ne, while spin entanglements in other nuclei are left for future discussions.
Several \twop-emitting nuclides are possibly generated in astrophysical processes \cite{1995Gorres,2001Schatz,13Olsen}.
If \twop-spin correlations exist in these nuclides, that indicates ``natural entanglement'', which emerges independently of artificial processes.
This is in contrast to that known entangled states have been artificially produced \cite{1978Clauser,1982Aspect,1998Weihs,2000JianWei,2001Rowe,2002Barrett,2010Scheidl,2015Giustina,2015Shalm,2017Vasilyev,2023Storz}.

\appendix
We thank Hideyuki Sakai, Masaki Sasano, Simin Wang, and Tokurou Fukui for discussions.
Following programs are acknowledged:
(i) Multi-disciplinary Cooperative Research Program (MCRP) in FY2024 and FY2025 by Center for Computational Sciences, University of Tsukuba (project ID wo23i034);
(ii) cooperative project of Yukawa-21 in Yukawa Institute for Theoretical Physics, Kyoto University.
The data supporting the findings in this work are not publicly available.
Those will be available from the authors upon reasonable request.

%<>\bibliographystyle{elsarticle}
%<>\bibliography{zb_full.bib}

\begin{thebibliography}{10}
\expandafter\ifx\csname url\endcsname\relax
  \def\url#1{\texttt{#1}}\fi
\expandafter\ifx\csname urlprefix\endcsname\relax\def\urlprefix{URL }\fi
\expandafter\ifx\csname href\endcsname\relax
  \def\href#1#2{#2} \def\path#1{#1}\fi

\bibitem{04Bell}
J.~S. Bell, Speakable and Unspeakable in Quantum Mechanics, 2nd Edition,
  Collected Papers on Quantum Philosophy, Cambridge University Press,
  Cambridge, UK, 2004.

\bibitem{64Bell}
J.~Bell, Physics 1 (1964) 195.

\bibitem{1969CHSH}
J.~F. Clauser, M.~A. Horne, A.~Shimony, R.~A. Holt, Phys. Rev. Lett. 23 (1969)
  880--884.

\bibitem{1982Aspect}
A.~Aspect, P.~Grangier, G.~Roger, Phys. Rev. Lett. 49 (1982) 91--94.

\bibitem{1998Weihs}
G.~Weihs, T.~Jennewein, C.~Simon, H.~Weinfurter, A.~Zeilinger, Phys. Rev. Lett.
  81 (1998) 5039--5043.

\bibitem{2001Rowe}
M.~A. Rowe, D.~Kielpinski, V.~Meyer, C.~A. Sackett, W.~M. Itano, C.~Monroe,
  D.~J. Wineland, Nature 409 (2001) 791--794.

\bibitem{2015Giustina}
M.~Giustina, M.~A.~M. Versteegh, S.~Wengerowsky, J.~Handsteiner, A.~Hochrainer,
  K.~Phelan, F.~Steinlechner, J.~Kofler, J.-A. Larsson, C.~Abell\'an, W.~Amaya,
  V.~Pruneri, M.~W. Mitchell, J.~Beyer, T.~Gerrits, A.~E. Lita, L.~K. Shalm,
  S.~W. Nam, T.~Scheidl, R.~Ursin, B.~Wittmann, A.~Zeilinger, Phys. Rev. Lett.
  115 (2015) 250401.

\bibitem{2015Shalm}
L.~K. Shalm, E.~Meyer-Scott, B.~G. Christensen, P.~Bierhorst, M.~A. Wayne,
  M.~J. Stevens, T.~Gerrits, S.~Glancy, D.~R. Hamel, M.~S. Allman, K.~J.
  Coakley, S.~D. Dyer, C.~Hodge, A.~E. Lita, V.~B. Verma, C.~Lambrocco,
  E.~Tortorici, A.~L. Migdall, Y.~Zhang, D.~R. Kumor, W.~H. Farr, F.~Marsili,
  M.~D. Shaw, J.~A. Stern, C.~Abell\'an, W.~Amaya, V.~Pruneri, T.~Jennewein,
  M.~W. Mitchell, P.~G. Kwiat, J.~C. Bienfang, R.~P. Mirin, E.~Knill, S.~W.
  Nam, Phys. Rev. Lett. 115 (2015) 250402.

\bibitem{2006Sakai}
H.~Sakai, T.~Saito, T.~Ikeda, K.~Itoh, T.~Kawabata, H.~Kuboki, Y.~Maeda,
  N.~Matsui, C.~Rangacharyulu, M.~Sasano, Y.~Satou, K.~Sekiguchi, K.~Suda,
  A.~Tamii, T.~Uesaka, K.~Yako, Phys. Rev. Lett. 97 (2006) 150405.

\bibitem{1976Rachti}
M.~Lamehi-Rachti, W.~Mittig, Phys. Rev. D 14 (1976) 2543--2555.

\bibitem{2004Polachic}
C.~Polachic, C.~Rangacharyulu, A.~{van den Berg}, S.~Hamieh, M.~Harakeh,
  M.~Hunyadi, M.~{de Huu}, H.~W\"{o}rtche, J.~Heyse, C.~B\"{a}umer, D.~Frekers,
  S.~Rakers, J.~Brooke, P.~Busch, Physics Letters A 323~(3) (2004) 176--181.

\bibitem{02Gri_15}
L.~Grigorenko, R.~Johnson, I.~Mukha, I.~Thompson, M.~Zhukov, The European
  Physical Journal A 15 (2002) 125--129.

\bibitem{02Pfu}
M.~Pftzner, E.~Badura, C.~Bingham, B.~Blank, M.~Chartier, H.~Geissel,
  J.~Giovinazzo, L.~Grigorenko, R.~Grzywacz, M.~Hellstrm, Z.~Janas,
  J.~Kurcewicz, A.~Lalleman, C.~Mazzocchi, I.~Mukha, G.~Mnzenberg, C.~Plettner,
  E.~Roeckl, K.~Rykaczewski, K.~Schmidt, R.~Simon, M.~Stanoiu, J.-C. Thomas,
  The European Physical Journal A - Hadrons and Nuclei 14~(3) (2002) 279--285.

\bibitem{2012Pfu_rev}
M.~Pf\"utzner, M.~Karny, L.~V. Grigorenko, K.~Riisager, Rev. Mod. Phys. 84
  (2012) 567--619.

\bibitem{2009Gri_rev}
L.~V. Grigorenko, Physics of Particles and Nuclei 40~(5) (2009) 674--714.

\bibitem{2016Golubkova}
T.~Golubkova, X.-D. Xu, L.~Grigorenko, I.~Mukha, C.~Scheidenberger, M.~Zhukov,
  Physics Letters B 762 (2016) 263--270.

\bibitem{08Blank_01}
B.~Blank, M.~Ploszajczak, Reports on Progress in Physics 71~(4) (2008) 046301.

\bibitem{08Blank_02}
B.~Blank, M.~Borge, Progress in Particle and Nuclear Physics 60~(2) (2008) 403
  -- 483.

\bibitem{2023Pfutzner_rev}
M.~Pf\"{u}tzner, I.~Mukha, S.~Wang, Progress in Particle and Nuclear Physics
  132 (2023) 104050.

\bibitem{2019Qi_rev}
C.~Qi, R.~Liotta, R.~Wyss, Progress in Particle and Nuclear Physics 105 (2019)
  214--251.

\bibitem{2003Dean}
D.~J. Dean, M.~Hjorth-Jensen, Rev. Mod. Phys. 75~(2) (2003) 607--656.

\bibitem{05BB}
D.~Brink, R.~Broglia, Nuclear Superfluidity: Pairing in Finite Systems,
  Cambridge Monographs on Particle Physics, Nuclear Physics and Cosmology,
  Cambridge University Press, Cambridge, UK, 2005.


\bibitem{2006Matsuo}
M.~Matsuo, Phys. Rev. C 73 (2006) 044309.

\bibitem{07Hagi_01}
K.~Hagino, H.~Sagawa, J.~Carbonell, P.~Schuck, Phys. Rev. Lett. 99 (2007)
  022506.

\bibitem{2010Oishi}
T.~Oishi, K.~Hagino, H.~Sagawa, Phys. Rev. C 82 (2010) 024315.

\bibitem{2014Oishi}
T.~Oishi, K.~Hagino, H.~Sagawa, Phys. Rev. C 90 (2014) 034303.

\bibitem{2017Oishi}
T.~Oishi, M.~Kortelainen, A.~Pastore, Phys. Rev. C 96 (2017) 044327.

\bibitem{2025Oishi_16Ne}
T.~Oishi, M.~Kimura, Phys. Rev. C 111 (2025) 044319.

\bibitem{2025Oishi_QUBITE}
T.~Oishi, Physics Letters B 862 (2025) 139361.

\bibitem{2021Wang_Naza}
S.~M. Wang, W.~Nazarewicz, Phys. Rev. Lett. 126 (2021) 142501.

\bibitem{2014Brown}
K.~W. Brown, R.~J. Charity, L.~G. Sobotka, Z.~Chajecki, L.~V. Grigorenko, I.~A.
  Egorova, Y.~L. Parfenova, M.~V. Zhukov, S.~Bedoor, W.~W. Buhro, J.~M. Elson,
  W.~G. Lynch, J.~Manfredi, D.~G. McNeel, W.~Reviol, R.~Shane, R.~H. Showalter,
  M.~B. Tsang, J.~R. Winkelbauer, A.~H. Wuosmaa, Phys. Rev. Lett. 113 (2014)
  232501.

\bibitem{2016Charity_EPJCON}
R.~J. Charity, The European Physical Journal Web of Conferences 117 (2016)
  06001.

\bibitem{12Gri}
L.~V. Grigorenko, I.~A. Egorova, R.~J. Charity, M.~V. Zhukov, Phys. Rev. C 86
  (2012) 061602.

\bibitem{1980Tsirelson}
B.~S. Cirel'son, Letters in Mathematical Physics 4 (1980) 93--100.

\bibitem{NNDC_Chart}
N.~N. D.~C. Brookhaven National~Laboratory, Chart of nuclides in nudat 3.0,
  https://www.nndc.bnl.gov/nudat3/ (2022).


\bibitem{2021Wang_AME}
M.~Wang, W.~Huang, F.~Kondev, G.~Audi, S.~Naimi, Chinese Physics C 45~(3)
  (2021) 030003.

\bibitem{02Gri}
L.~V. Grigorenko, I.~G. Mukha, I.~J. Thompson, M.~V. Zhukov, Phys. Rev. Lett.
  88 (2002) 042502.

\bibitem{08Muk}
I.~Mukha, L.~Grigorenko, K.~S\"ummerer, L.~Acosta, M.~A.~G. Alvarez,
  E.~Casarejos, A.~Chatillon, D.~Cortina-Gil, J.~M. Espino, A.~Fomichev, J.~E.
  Garcia-Ramos, H.~Geissel, J.~G\'omez-Camacho, J.~Hofmann, O.~Kiselev,
  A.~Korsheninnikov, N.~Kurz, Y.~Litvinov, I.~Martel, C.~Nociforo, W.~Ott,
  M.~Pf\"utzner, C.~Rodriguez-Tajes, E.~Roeckl, M.~Stanoiu, H.~Weick, P.~J.
  Woods, Phys. Rev. C 77 (2008) 061303.

\bibitem{10Muk}
I.~Mukha, K.~S\"ummerer, L.~Acosta, M.~A.~G. Alvarez, E.~Casarejos,
  A.~Chatillon, D.~Cortina-Gil, I.~A. Egorova, J.~M. Espino, A.~Fomichev, J.~E.
  Garcia-Ramos, H.~Geissel, J.~G\'omez-Camacho, L.~Grigorenko, J.~Hofmann,
  O.~Kiselev, A.~Korsheninnikov, N.~Kurz, Y.~A. Litvinov, E.~Litvinova,
  I.~Martel, C.~Nociforo, W.~Ott, M.~Pf\"utzner, C.~Rodriguez-Tajes, E.~Roeckl,
  M.~Stanoiu, N.~K. Timofeyuk, H.~Weick, P.~J. Woods, Phys. Rev. C 82 (2010)
  054315.

\bibitem{77Thom}
D.~Thompson, M.~Lemere, Y.~Tang, Nuclear Physics A 286~(1) (1977) 53 -- 66.

\bibitem{2003Peters}
W.~A. Peters, T.~Baumann, D.~Bazin, B.~A. Brown, R.~R.~C. Clement, N.~Frank,
  P.~Heckman, B.~A. Luther, F.~Nunes, J.~Seitz, A.~Stolz, M.~Thoennessen,
  E.~Tryggestad, Phys. Rev. C 68 (2003) 034607.

\bibitem{1995Gorres}
J.~G\"{o}rres, M.~Wiescher, F.-K. Thielemann, Phys. Rev. C 51 (1995) 392--400.

\bibitem{2001Schatz}
H.~Schatz, A.~Aprahamian, V.~Barnard, L.~Bildsten, A.~Cumming, M.~Ouellette,
  T.~Rauscher, F.-K. Thielemann, M.~Wiescher, Phys. Rev. Lett. 86 (2001)
  3471--3474.

\bibitem{13Olsen}
E.~Olsen, M.~Pf\"{u}tzner, N.~Birge, M.~Brown, W.~Nazarewicz, A.~Perhac, Phys.
  Rev. Lett. 110 (2013) 222501.

\bibitem{1978Clauser}
J.~F. Clauser, A.~Shimony, Reports on Progress in Physics 41~(12) (1978) 1881.

\bibitem{2000JianWei}
J.-W. Pan, D.~Bouwmeester, M.~Daniell, H.~Weinfurter, A.~Zeilinger, Nature 403
  (2000) 515--519.

\bibitem{2002Barrett}
J.~Barrett, D.~Collins, L.~Hardy, A.~Kent, S.~Popescu, Phys. Rev. A 66 (2002)
  042111.

\bibitem{2010Scheidl}
T.~Scheidl, R.~Ursin, J.~Kofler, S.~Ramelow, X.-S. Ma, T.~Herbst,
  L.~Ratschbacher, A.~Fedrizzi, N.~K. Langford, T.~Jennewein, A.~Zeilinger,
  Proceedings of the National Academy of Sciences 107~(46) (2010) 19708--19713.

\bibitem{2017Vasilyev}
D.~Vasilyev, F.~O. Schumann, F.~Giebels, H.~Gollisch, J.~Kirschner, R.~Feder,
  Phys. Rev. B 95 (2017) 115134.

\bibitem{2023Storz}
S.~Storz, J.~Sch\"{a}r, A.~Kulikov, P.~Magnard, P.~Kurpiers, J.~L\"{u}tolf,
  T.~Walter, A.~Copetudo, K.~Reuer, A.~Akin, J.-C. Besse, M.~Gabureac, G.~J.
  Norris, A.~Rosario, F.~Martin, J.~Martinez, W.~Amaya, M.~W. Mitchell,
  C.~Abellan, J.-D. Bancal, N.~Sangouard, B.~Royer, A.~Blais, A.~Wallraff,
  Nature 617 (2023) 265--270.

\end{thebibliography}

%\begin{thebibliography}{00}
%%\bibitem{label}
%%Text of bibliographic item
%\bibitem{}
%\end{thebibliography}

\end{document}